\documentclass{ws-procs11x85}
\usepackage{ws-procs-thm}       
\usepackage[utf8]{inputenc}
\usepackage[T1]{fontenc}

\begin{document}

\title{Learning Contextual Hierarchical Structure of Medical Concepts with Poincair\'e Embeddings to Clarify Phenotypes}
\author{Brett K. Beaulieu-Jones, Isaac S. Kohane and Andrew L. Beam$^\dag$}

\address{Department of Biomedical Informatics, Harvard Medical School,\\
Boston, MA 02115, USA\\
$^\dag$E-mail: Andrew\_Beam@hms.harvard.edu\\
dbmi.hms.harvard.edu}

\begin{abstract}
Biomedical association studies are increasingly done using clinical concepts, and in particular diagnostic codes from clinical data repositories as phenotypes. Clinical concepts can be represented in a meaningful, vector space using word embedding models. These embeddings allow for comparison between clinical concepts or for straightforward input to machine learning models. Using traditional approaches, good representations require high dimensionality, making downstream tasks such as visualization more difficult. We applied Poincar\'e embeddings in a 2-dimensional hyperbolic space to a large-scale administrative claims database and show performance comparable to 100-dimensional embeddings in a euclidean space. We then examine disease relationships under different disease contexts to better understand potential phenotypes.
\end{abstract}


\copyrightinfo{\copyright\ 2018 The Authors. Open Access chapter published by World Scientific Publishing Company and distributed under the terms of the Creative Commons Attribution Non-Commercial (CC BY-NC) 4.0 License.}

\bodymatter

\section{Introduction}\label{aba:Introduction}
Word embeddings\cite{Mikolov2013EfficientSpace} are a popular way to represent natural language and have seen wide use in machine learning applied to document classification\cite{Kusner2015FromDistances, Yang2016HierarchicalClassification}, machine translation\cite{Cho2014LearningTranslation, Bahdanau2014NeuralTranslate}, sentiment analysis\cite{AssociationforComputationalLinguistics.Meeting45th:2007:Prague2007ACLRepublic.}, and question answering\cite{Zhou2015PredictingModel,Bordes2014OpenModels}. Clinical concept embeddings extend this approach to model healthcare events \cite{beam2018clinical,Choi2016LearningConcepts.,Ching2018OpportunitiesMedicine,Beaulieu-Jones2018MachineData}, and have been particularly useful modeling longitudinal clinical data\cite{ChoiDoctorNetworks,Lipton2015LearningNetworks,Rajkomar2018ScalableRecords,Beaulieu-Jones2018MappingLicense}. Traditional approaches such as word2vec\cite{Mikolov2013EfficientSpace} and GloVe \cite{PenningtonGlove:Representation} embed entities within a Euclidean space. 

However, recent work by Nickel and Kiela on \emph{Poincar\'e embeddings}\cite{NickelPoincareRepresentations} claims to provide better embedding representations of hierarchically structured data using a hyperbolic embedding space within the Poincar\'e ball. This n-dimensional hyperbolic space has a significantly higher capacity than the Euclidean space, which allows it to effectively embed structured trees while preserving distance relationships\cite{Gromov.1987HyperbolicGroups.,BonnabelStochasticManifolds,BordesTranslatingData,Krioukov2010HyperbolicNetworks}. Moreover, this space allows for embedding of hierarchical, tree-like structures, as Nickel and Kiela \cite{NickelPoincareRepresentations} observed high fidelity embeddings of ontologies. This has an obvious relevance to medical concepts, given many have an inherent tree structure (e.g. disease nosology) that should be recapitulated in the embedding space. 

When clinicians consider a disease, they examine the disease in the context of the patient's overall environment\cite{Beaulieu-Jones2016Semi-supervisedStratification}. For example, renal failure caused by poor blood flow to the kidneys as a result of long-term hypertension would be considered differently from renal failure as the result of a specific infection or immune system disorder like Lupus\cite{Salem2002PathophysiologyFailure.}. Accurate and precise phenotyping is critical to modern clinical studies using the electronic healthcare record (EHR) and other '-omic' associations studies (e.g. genomic, transcriptomic, metabolomic). Misclassified phenotypes have a severe effect on tests of association and require increased sample sizes to maintain constant power\cite{Smith2013GenomeResponses.,Buyske2009WhenTrios,Rekaya2016AnalysisStudies.}. Increases in genetic testing and the availability of clinical data repositories (Electronic Health Record, Administrative Claims, large-scale Cohort) have enabled PheWAS association studies to be performed without the need to target and recruit specific populations for each individual study\cite{Verma2018PheWASGeisinger.,Denny2010PheWAS:Associations,Pendergrass2013Phenome-WideNetwork}. It is important to develop methods that enable researchers to consider a specific disease or phenotype in the context of the overall patient and environment.

We applied Poincar\'e embeddings to a large-scale administrative claims database to examine how the relationships of different conditions changed in distinct contexts. 
Our hypothesis was that the increased representational capacity offered by Poincar\'e embeddings and their ability to naturally model hierarchical data would result in improved embeddings for clinical concepts. We first demonstrate this by showing they can accurately reconstruct the ICD-9 hierarchy on synthetic data. Next we show that they find an improved representation on real data relative to traditional embedding approaches at the same number of dimensions. We conclude with a disease-specific embedding hierarchy within an obese population. Our results could provide a better representation of disease and allow for more accurate machine learning models as well as the fine-tuning of targeted phenotypes for association studies.

\section{Methods}
To examine the effectiveness of Poincar\'e embeddings for clinical concept embedding, we: 1.) trained Poincar\'e embeddings on the ICD-9 hierarchy as validation of parent-child tuples, 2a.)  selected and preprocessed chronological member sequences of each diagnosis experienced for a specified cohort (e.g. obese vs. no metabolic disorders diagnosed), 2b.) Learned distributed vector representations for the real data by training a Poincar\'e embedding model in a two-dimensional space. 3.) Visualized the Poincar\'e embeddings in a two dimensional space. 4a.) Constructed a distance matrix within the hyperbolic space. 4b.) Analyzed the distance matrix to measure how effectively the embeddings represent clinical groupings (e.g. ICD9 Chapter, Sub-chapter and major codes).

\subsection{Source Code}\label{aba:SourceCode}
The source code used for the analyses in this work are freely available on Github  (https://github.com/brettbj/poincareembeddings) under a permissive open source license. The optimized C++ Poincare Embedding implementation by Tatsuya Shirakawa is available under the MIT license (https://github.com/TatsuyaShirakawa/poincare-embedding).

\subsection{Data Source}\label{aba:DataSource}
These analyses were performed using de-identified insurance administration data including diagnostic billing codes from January 1, 2008 until February 29, 2016 for more than 63 million members. The database does not include any socioeconomic, race or ethnicity data. The Institutional Review Board at Harvard Medical School waived the requirement for approval as it deemed analyses of the de-identified dataset to be non-human subjects research.

The data to rebuild the reference ICD9 hierarchy tree is available in the GitHub repository (https:/github.com/brettbj/poincareembeddings/data/icd9.tsv).

\subsection{Data Selection and Preprocessing
}\label{aba:Preprocessing}
\subsubsection{Reference ICD9 Example}\label{aba:referenceICD9}
We first benchmarked against a known hierarchy, the ICD9 2015-Clinical Modification code ontology. To do this we extracted the ICD9 codes into four levels: 1.) Chapters (e.g. codes 390-459: Diseases of the circulatory system), 2.) Sub-chapters (e.g. codes 401-405: Hypertensive disease), 3.) Major Codes (e.g. code 401: Essential hypertension), and 4.) Detail level codes (e.g. code 401.0: Hypertension, malignant). We assigned relationships between each detail level code and the chapter, sub-chapter and major code it belonged to, each major code to the appropriate sub-chapter and chapter, and each sub-chapter to the chapter it belonged to.

\subsubsection{Real Member Analyses}
We performed cohort analyses by defining different study groups. First we included ten million randomly selected members (without replacement) who were enrolled for at least two years from the database of 63 million members. Next we separated two groups based on obesity diagnoses: 1.) ten million members who do not have a diagnosis for metabolic disorders with ICD9 codes between 270 and 279 2.) 3.38 million members who were diagnosed with obesity ICD9 codes (278.00 and 278.01).

Poincar\'e embeddings learn distributed vector representations from hierarchical data (e.g. a directed graph or tree). The input to the model is a list of tuples of the form $<A,B>$, which indicates that $A$ and $B$ have some form of unspecified relationship (e.g. \emph{parent of}, \emph{co-occurs with}, etc). In our case, the list of relationships specify that two diagnoses occurred sequentially, within a one year period, and had to occur more than ten total times and in more than 2\% of all diagnoses.

\subsection{Poincar\'e Embeddings}\label{aba:Poincare}
The key way in which Poincar\'e embeddings differ from traditional approaches is the distance metric which is used to compare the embeddings for two concepts. This distance metric is given in equation \ref{aba:eq1}:

\begin{equation}
dist((x_1, y_1), (x_2, y_2)) = arccosh(1 + \frac{(x_2-x_1)^2 + (y_2-y_1)^2}{2y_1y_2})
\label{aba:eq1}
\end{equation}

Equation \ref{aba:eq1} shows the distance between two points in the Poincar\'e ball hyperbolic space.
\newline

Training a Poincar\'e embedding model occurs by maximizing the distance (Equation \ref{aba:eq1}) between unconnected nodes or diagnoses while minimizing the distance between highly connected nodes. This is done using a stochastic Riemannian optimization method, specifically stochastic gradient descent on riemmanian manifolds as seen in Bonnabel\cite{BonnabelStochasticManifolds}.

\subsection{Processing and Evaluating Embeddings}\label{aba:Processing}

Once each concept is embedded into a two dimensional space, it is possible to calculate the pair-wise distance between all concepts using Equation \ref{aba:eq1}. To assess how well the embeddings captured the ICD hierarchy on real data, we compared the average distances between concepts in the same ICD9 major code, sub-chapter and chapter against the distances of all other concepts. We then compared the capacity of a two-dimensional Poincair\'e space with varying size euclidean spaces. To do this, we repeated distance calculations with the clinical concept embeddings trained in a euclidean space on more than 63 million members in 2, 10 and 100 dimensions from Beam et al.\cite{beam2018clinical} To normalize the distance comparisons between hyperbolic and euclidean spaces, we compared the ratio of distances between ICD codes within the same major, sub-chapter and chapter and the other ICD codes outside of the major, sub-chapter, and chapter.

\section{Results}\label{aba:Results}

\subsection{ICD9 Hierarchy Evaluation }

\begin{figure}
	\includegraphics[width=\linewidth]{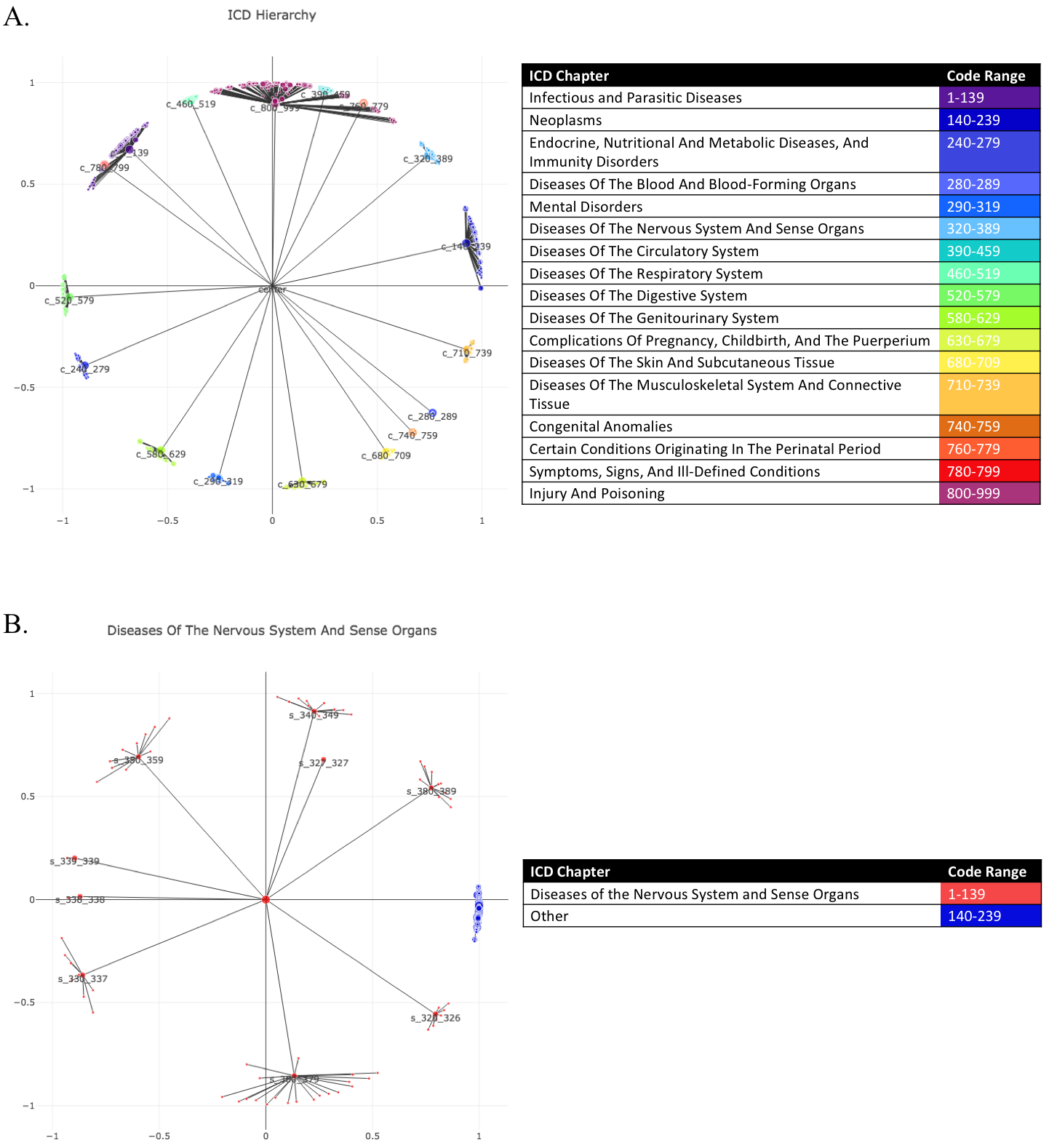}
	\caption{ICD Example All codes}
	\label{fig:ICD Tree}
\end{figure}

To evaluate the method with a known ground truth, we embedded the ICD9 hierarchy and then reconstructed it as a tree. Because there are no counts included, stochasticity for all relationships at the same level (Chapter, Sub-chapter, Major, Detail) was expected. Figure \ref{fig:ICD Tree} shows the reconstructed tree of the predefined ICD9 tree. This served as evidence that Poincair\'e embeddings can effectively embed a clean ICD9 hierarchy.

\subsection{Poincar\'e Embeddings on 10 Million Members}
We then trained Poincar\'e embeddings in a two-dimensional space for 10 million randomly selected members (Table 1). 

\begin{table}[]
    \centerline{Table 1 Member Demographics of the Training Data}
    \centering
    \begin{tabular}{l|l}
    \multicolumn{2}{ c }{Demographics} \\ 
    \hline 
    Male   					 & 40.4\%        \\
    Female					 & 59.6\%        \\
    Age (2016)               & 48.66 (22.68) \\
    ICD9 Diagnoses        & 22.38 (28.70) \\	
    \end{tabular}\par
  \label{tab:Demographics}
\end{table}

\begin{figure}
	\centering
	\includegraphics[width=0.8\linewidth]{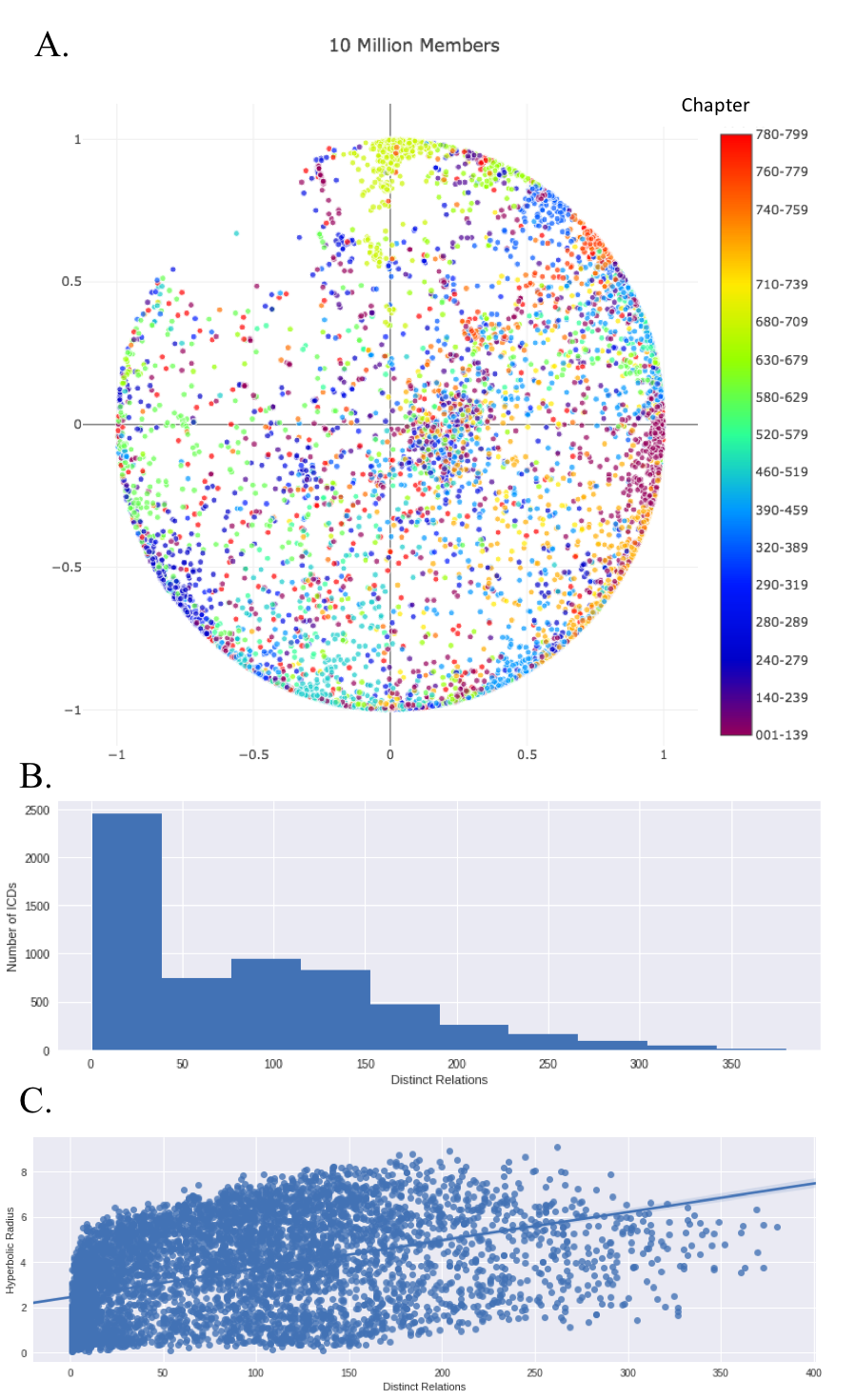}
	\caption{A.) ICD9 Diagnoses Codes Embedded in a two-dimensional space. B.) Examination of the number of distinct relations for each ICD9 code. C.) Examination of the Correlation between the number of distinct relations and hyperbolic distance.}
	\label{fig:GenEmbed}
\end{figure}

Figure \ref{fig:GenEmbed}A shows the ICD9 concepts (labeled by chapter) embedded in a two-dimensional space. While there were over 223 million total diagnoses, the majority of concepts had less than 100 distinct relations (Figure \ref{fig:GenEmbed}B) and the number of distinct relations was correlated with the distance from the origin ($R^2 = 0.61$) (Figure \ref{fig:GenEmbed}C). 

Figure \ref{fig:GenEmbed} shows that the ICD hierarchy is correctly reconstructed using by the Poincar\'e embeddings in two dimensions. The distances between ICD codes in the same major, sub-chapter and chapter are smaller than the distances across different major codes, sub-chapters and chapters (Table 2). This shows that Poincar\'e embeddings are representing the data in a way that has similarities with the human-defined ICD9 hierarchy. 

\begin{table}
\centerline{Table 2. Hyperbolic Distance comparison within Major, Sub-chapter and Chapter}
\centering
\begin{tabular}{l|l|l}
Category    & In Category & Outside of Category   \\ \hline
Major       & 3.87 (1.71) & 5.89 (1.92)           \\
Sub-chapter & 4.47 (1.73) & 5.89 (1.92)           \\
Chapter     & 4.91 (1.81) & 5.91 (1.94)           
\label{tab:groupDistances}
\end{tabular}
\end{table}

\subsection{Comparison with Euclidean Embeddings}

To evaluate Poincar\'e embeddings against traditional euclidean embeddings, we compared the 2-dimensional Poincar\'e embeddings with 2, 10 and 100 dimension embeddings. The Poincar\'e embeddings were trained on 10 million randomly selected members. Running the preprocessing pipeline required 42 minutes on 16 cores but training the embeddings required only 49 seconds on 16 cores. All euclidean embeddings were trained on more than 63 million members. Table 3 shows the ratios of the mean distances of ICD codes in the same category over ICD codes in all other categories. We show the ratio to allow for comparison between Poincar\'e and Euclidean distances. As the dimensionality of the euclidean embeddings increased, the ratio of distance in-group vs. out of group decreased, indicating that the higher capacity enabled a better representation. The 2-dimensional Poincar\'e embeddings compared most closely to the 100-dimensional euclidean embeddings.

\begin{table}[]
\centering{Table 3 Distance (ratio) comparison between Poincar\'e (2-dimensional) and Euclidean (2, 10, \& 100-dimensional) within Major, Sub-chapter and Chapter.}
\begin{tabular}{l|l|l|l|l}
Category    & Poincaire (2d) & Euclidean (2d) & Euclidean (10d) & Euclidean (100d) \\ \hline
Major       & 0.657          & 0.758         & 0.668          & 0.649           \\
Sub-chapter & 0.759          & 0.863         & 0.794          & 0.774           \\
Chapter     & 0.831          & 0.894         & 0.856          & 0.830          
\label{tab:distanceComparison}
\end{tabular}\par
\end{table}

\begin{figure}
	\centering
	\includegraphics[width=\linewidth]{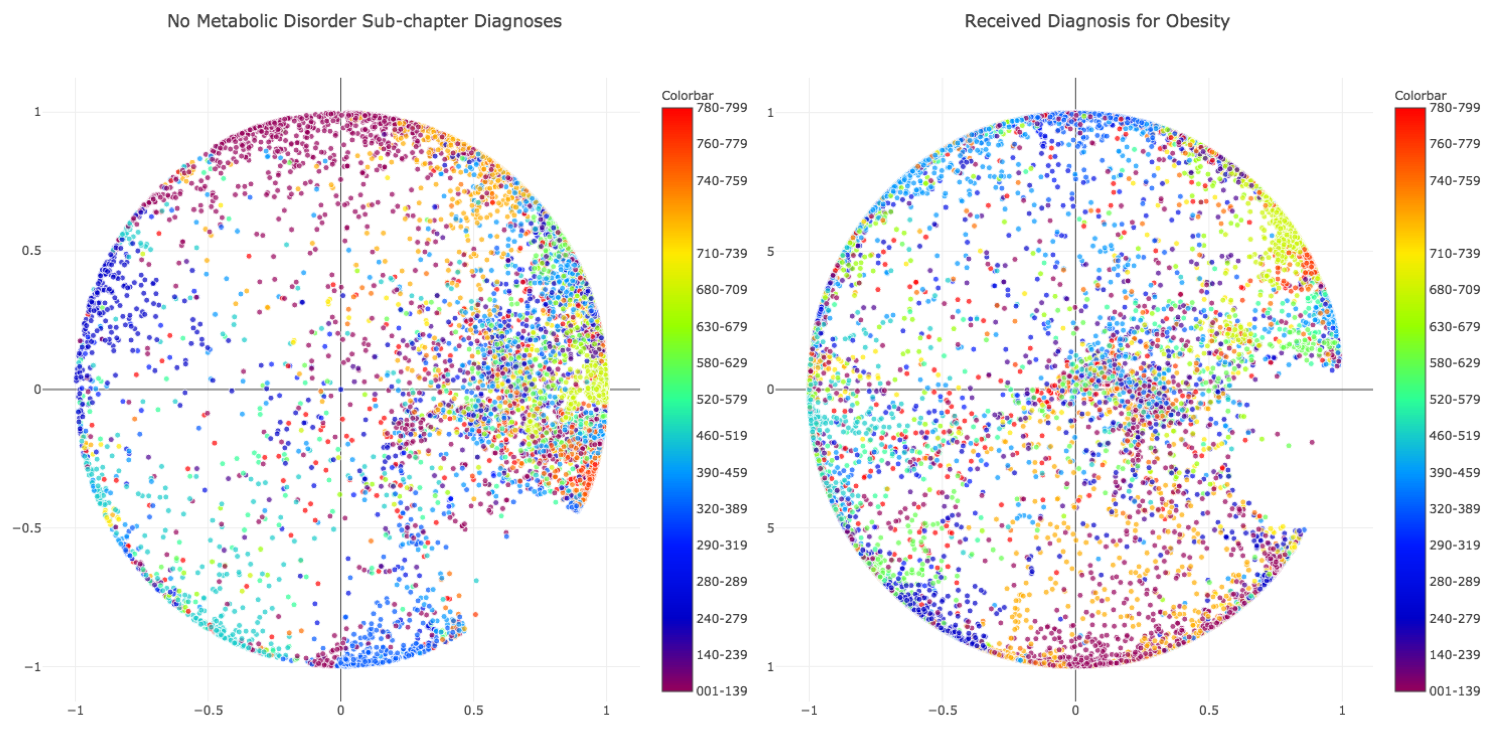}
	\caption{A.) Poincar\'e Embeddings trained on 10M members with no metabolic disorder diagnoses (centered on type 2 diabetes). B.) Poincar\'e Embeddings trained on 3.38M members diagnosed with obesity (centered on type 2 diabetes).}
	\label{fig:ObeseCompare}\par
\end{figure}

\subsection{Cohort Specific Embeddings}
Finally, we trained two separate Poincar\'e embeddings on patients with either: 1.) no prior diagnoses from the sub-chapter of metabolic disorders between ICD code 270 and 279 (N=10,000,000) and 2.) members diagnosed with obesity (ICD codes 278.00, 278.01, N=3,377,267) to first visualize the differences in the context of type 2 diabetes mellitus (Figure \ref{fig:ObeseCompare}). Because the Poincar\'e embedding model was trained in 2-dimensions this was done without any further dimensionality reduction step. 

\begin{table}[]
  \centering{Table 4. ICD9 Codes with the largest changes in distance from Type 2 Diabetes (250.00).}
  \begin{tabular}{l|l|l}
    & ICD    & Description \\ \hline
  1 & 553.21 & Incisional hernia               \\
  2 & 786.09 & Other Respiratory Abnormalities \\
  3 & 599.0  & Urinary tract infection 		 \\
  4 & 285.9  & Anemia				             \\
  5 & 571    & Chronic Liver Disease           \\
  6 & 583.6  & Nephritis			             \\
  7 & 724.5 & Backache, unspecified			 \\
  8 & 710.5 & Eosinophilia myalgia syndrome   \\
  9 & 796.2 & Elevated blood pressure w/o hypertension        \\
  10 & 719.46 & Pain in Leg                    
  \end{tabular}\par
  \label{tab:t2dMovement}
\end{table}

We then examined the diseases in the closest quartile of either cohort to determine which showed the greatest movement from type 2 diabetes (Table 4). Of note, 22 of the top 50 were pain related and there are numerous links in the literature between both obesity (particularly joint and fibromyalgia \cite{Okifuji2015TheObesity.,McVinnie2013ObesityPain.}) and type 2 diabetes (particularly neuropathy \cite{Young1993APopulation}) with pain. 

\section{Discussion and Conclusion}\label{aba:Discussion}
Machine learning has great potential to improve the delivery of healthcare to patients, but many methodological challenges remain before this potential can be realized \cite{Beam2018BigCare, Ghassemi2018OpportunitiesHealthcare}. In this work, we showed the increased capacity and hierarchical positioning of Poincar\'e embedding models can be useful to learn representations of disease diagnosis codes. Two-dimensional Poincar\'e embeddings were on par with 100-dimension euclidean embeddings when compared to the human-defined ICD hierarchy. Importantly the extra capacity of Poincar\'e embeddings may directly allow for visualization in a two-dimensional space, while traditional euclidean embedding techniques require an additional dimensionality reduction step (PCA, t-SNE, UMAP). Many of these techniques are non-deterministic and may not preserve global structure. 

An important limitation of our current method is that the pre-processing step constructs binary relations between concepts whenever they occur with a specified threshold (more than 10 occurrences and  2\% of cases). It is likely that additional information could be learned by encoding the actual frequency between concepts. In addition, it could be useful to evaluate additional distance matrices that have worked well for hierarchical problems in other domains, such as pg-gram and Edit distance \cite{HassanComparisonDatabases}.

There are significant opportunities to expand on and apply these techniques to biomedical domains in order to examine and consider phenotypic context when performing associations. We are especially interested in the ability to contextualize a phenotype for association studies by considering the way ICD code relationships change given comorbidities. For example, start by measuring the way Poincar\'e embeddings change given a comorbidity (e.g. type 2 diabetes given metabolic disorder). If there are significant changes, it may be helpful to design association studies to separate endpoints, for example diabetes with no prior metabolic disorders and diabetes with prior metabolic disorders. In this case, the disease etiology may be distinct, and therefore we would expect the potential for different genetic drivers.

\section{Acknowledgments}
The authors thank Tatsuya Shirakawa for developing and open-sourcing an efficient implementation of the Poincar\'e Embedding Model. This work was supported in part by NLM grant 4 T15 LM007092-25. 

\bibliographystyle{ws-procs11x85}
\bibliography{Mendeley}

\end{document}